# Versatile entropic measure of grey level inhomogeneity

Ryszard Piasecki[*]

*Faculty of Chemistry, University of Opole, Oleska 48, 45-052 Opole, Poland*

**Abstract**
The entropic measure for analysis of grey level inhomogeneity (GLI) is proposed as a function of length scale. It allows us to quantify the statistical dissimilarity of the actual macrostate and the maximizing entropy of the reference one. The maximums (minimums) of the measure indicate those scales at which higher (lower) average grey level inhomogeneity appears compared to neighbour scales. Even a deeply hidden statistical grey level periodicity can be detected by the equally distant minimums of the measure. The striking effect of multiple intersecting curves (MIC) of the measure has been revealed for pairs of simulated patterns, which differ in shades of grey or symmetry properties, only. This indicates for a non-trivial dependence of the GLI on length scale. In turn for evolving photosphere granulation patterns, the stability in time of the first peak position has been found. Interestingly, at initial steps of the evolution clearly dominates the third peak. This indicates for a temporary grouping of granules at length scale that may belong to mesogranulation phenomenon. This behaviour has similarities with that reported by Consolini, Berrilli et al. (2003, 2005) for binarized granulation images of a different data set.

## 1. Introduction sensitive measures

The morphological features of complex materials are vital for modelling and predicting their macroscopic properties, for instance, effective conductivity [1-4]. For binary micrographs of such materials, the quantitative characterization of the spatial distribution of pixels, in some cases allows correlate their properties and internal structure attributes. As the white and black pixels are finite size objects (FSO), one can assume that the latter represent hard-core particles. When points approximate particles, the system can be described by a complete set of $n$-point correlation functions [1]. In practice, their calculation is rather difficult with the exception of two-point correlation function.

The much simpler statistical method of finding length scale depending different morphological features for disordered systems gives normalized information entropy [5] or cluster diversity and cluster entropy [6, 7]. The length scale is defined by a side size of sampling square cell. However, the measures employ a set of discrete probabilities, which statistical meaning weakens at large length scales [5] or given low/high lattice site occupation probability [6, 7]. The spatial inhomogeneity for FSO distributions can be also evaluated by simple *mathematical* statistical measure [8] or *physical* entropic measure [9] based mainly on the combinatorial approach. For $L \times L$ patterns partitioned in non-overlapping cells, the mathematical measure is limited to the length scales being integer divisors of $L$. Free of such limitations, the exact physical entropic measure even the subtle self-similarity traces in model surface microstructures can detect [10]; there in Eq. (3) the sign '+' before the second term should be replaced by opposite one '−'. In turn, when a sliding cell sampling (SCS) is used, all length scales are also allowed for the mathematical measure. The last two measures initially slightly differ while at the other scales they well correlate. Even for the assumed minimal number of sampled cells of order 1000, the range of appropriate length scales should be no greater than $L - 31$ according to a simple condition for a good cell statistics given in [8].

---

[*] Fax: +48 77 4527101.
  *E-mail address:* piaser@uni.opole.pl.

On the other hand, experimental data and results of theoretical simulations are frequently presented through grey scale or colour images. The latter can be also carefully converted into proper grey scale patterns. It is obvious that more complex data can be encoded with this type of patterns compared to binary ones. Thus, the revealing of any of length scale depending characteristic features of grey level patterns is of some importance. The proposed here grey level generalization is a natural *completion* of the binary entropic measure [9, 10]. Instead of the spatial degrees of freedom involved formerly, the generalized measure considers specific degrees of freedom. Those relate to possible distributions (under some conditions) of every cell sum of grey level values inside the corresponding cell at given length scale. Then two configurational macrostates, defined in the next section, are compared per cell: the obtained on the basis of existent pattern and the calculated reference one related to the most uniform distribution of grey level values. Thus, the present entropic measure quantifies a kind of average grey level inhomogeneity over a range of length scales.

## 2. The generalized entropic measure

Let us assume that the initial pattern is a representative one for the investigated physical system. Instead of its standard partitioning employed in [9, 10] here we use SCS method, see for instance [8]. For a given grey scale pattern of size $L \times L$ there is $\kappa(k) = [L - k + 1]^2$ allowed positions of the sliding cell of size $k \times k$. When the sampled cells are placed in a non-overlapping manner, the obtained in this way auxiliary pattern $L_a \times L_a$, where $L_a \equiv [L - k + 1] k$ reproduces the general structure of the initial one, cf. Fig. 1(b). The SCS procedure provides local sums $G_i(k)$ of grey level values for each sampled $i$th cell of the initial pattern. At each length scale $1 \leq k \leq L$ the only restriction for local sums is the natural constraint given by

$$\sum_{i=1}^{\kappa} G_i(k) = G(k), \tag{1}$$

where $G(k)$ stands for the length scale depending total sum of grey level values.

For a given configuration at scale $k$, the actual macrostate, AM($k$), can be defined by the corresponding set, $\{G_i(k)\}_{AM} \equiv (G_1(AM),\ldots, G_\kappa(AM))$. Inside every cell of size $k \times k$ the $k^2$ of its unit cells each of size $1 \times 1$ can be numbered in sequence $1, 2,\ldots, k^2$. As we take into account grey level values from the range (0-255) some of the unit cells can be occupied by zero valued grey level. Within the simplest approach we consider all possible order dependent partitions (allowing some of the parts to be zero) of obtained local sum $G_i(k)$ over $k^2$ unit cells inside $i$th cell for $i = 1, 2,\ldots, \kappa$. In mathematics it is sometimes referred to as a *weak composition* [11]. Then we calculate the number $\Omega_{gr}(k)$ of realizations of AM($k$), i.e. the number of the appropriate configurational microstates that is the product of the ways that each of sampled cells can be populated under above conditions

$$\Omega_{gr}(k, G) = \prod_{i=1}^{\kappa} \binom{G_i + k^2 - 1}{k^2 - 1}. \tag{2}$$

Now, we make use of a microcanonical entropy, $S_{gr}(k) = \ln \Omega_{gr}(k)$, where the Boltzmann constant is set $k_B = 1$ for convenience. It is worth to notice that the entropic approach has been already utilized to binary patterns [9, 10]. To obtain at given length scale $k$ the highest possible value of the entropy, $S_{gr, max}(k) = \ln \Omega_{gr, max}(k)$, we need a reference macrostate, RM($k$), ensuring the maximal number $\Omega_{gr, max}(k)$ of its realizations. It is described by the appropriate set, $\{G_i(k)\}_{RM} \equiv (G_1(RM),\ldots, G_\kappa(RM))$, where each pair $i \neq j$ of sampled cells fulfils the simple condition $|G_i(k) - G_j(k)| \leq 1$. Then, the maximal number of the proper microstates can be written as

$$\Omega_{gr,\max}(k,G) = \binom{G_0+k^2-1}{k^2-1}^{\kappa-R_0}\binom{G_0+k^2}{k^2-1}^{R_0}, \tag{3}$$

where $R_0(k) = G(k)\bmod \kappa$ and $G_0(k) = [G(k)-R_0(k)]/\kappa$. Every microstate of the $\Omega_{gr,\max}$ set represents a reference macrostate $\{G_i \in (G_0, G_0+1)\}_{RM}$ with $\kappa - R_0$ and $R_0$ number of cells with sums $G_0$ and $G_0+1$ of grey levels, respectively.

Our aim is to characterize the average relative grey level inhomogeneity (GLI), at every length scale $k$. Therefore we need to evaluate a *deviation* per cell of a given AM from the suitable RM, that is a difference of the corresponding entropies. It is given by

$$S_{gr,\Delta}(k) = [S_{gr,\max}(k;RM) - S_{gr}(k;AM)]/\kappa$$

$$= \frac{\left[(\kappa-R_0)\ln\binom{G_0+k^2-1}{k^2-1} + R_0\ln\binom{G_0+k^2}{k^2-1} - \sum_{i=1}^{\kappa}\ln\binom{G_i+k^2-1}{k^2-1}\right]}{(L-k+1)^2}. \tag{4}$$

According to above definition, at boundary length scales we have $S_{gr,\Delta}(k=1) = S_{gr,\Delta}(k=L) = 0$. The maximums (minimums) of the measure indicate those scales at which higher (lower) departure of AM from RM per cell appears compared to neighbour scales. In particular, at scale denoted as $k_1$ the first maximum is a sign of formation of grey level clusters with a *different* local average of grey level values in comparison to the total average of grey level values in a whole pattern. The length scale $k_1$ only approximates spatial size of the clusters but says nothing about their shapes. The additional maximums should be read as occurrence of groups of such grey level clusters. The minimums correspond to grey level arrangements with the local sums relatively close to the distinguished sums $G_0$ and $G_0+1$ of grey levels mentioned above. Thus, one can say that a pattern has grey level arrangements more spatially homogeneous on average at those scales. When all minimums are equally distant from each other then a kind of grey level periodicity of a pattern can be identified. To illustrate the basic properties of our entropic measure a few examples of simulated and experimental grey level patterns will be given in the next section.

## 3. Examples

To examine the validity of our method, we implement it to four pairs of simulated grey level patterns and series of Sun photosphere grey-scale frames evolving in time. We begin with a binary $83 \times 83$ pattern of approximately equal black and white areas, cf. Fig. 1(a) in Ref. [8]. Next, the black (white) areas are randomly modified to obtain grey level darker (lighter) subdomains. Their averages of grey level values equal to 97.5 (157.5) for the more contrasted L-pattern and 112.5 (142.5) for the less contrasted R-one, see the insets in Fig. 1. Notice, that for the L-subdomain averages differ by $\delta(L) = 157.5 - 97.5 = 60$ while for the R merely by $\delta(R) = 30$. Nevertheless, the total averages of grey level values, $\varphi(L) \cong 127.67$ and $\varphi(R) \cong 127.59$, are very close to each other. We underline that the investigated GLI measure is a discrete function of length scale $k$. So, the continuous line, thick for L and thin for R, should be treated in this and next figures as a visual guide.

In Fig. 1(a) for both GLI curves the first maximum appears at $k_1 = 6$ indicating an appearance of small clusters with cell averages of grey levels different in comparison to the total average in a whole pattern. Those clusters are prevalent on rings, which are less or more diffused, see the corresponding auxiliary pattern $L_a(k) \times L_a(k)$ with $L_a(6) = 468$ in Fig. 1(b).

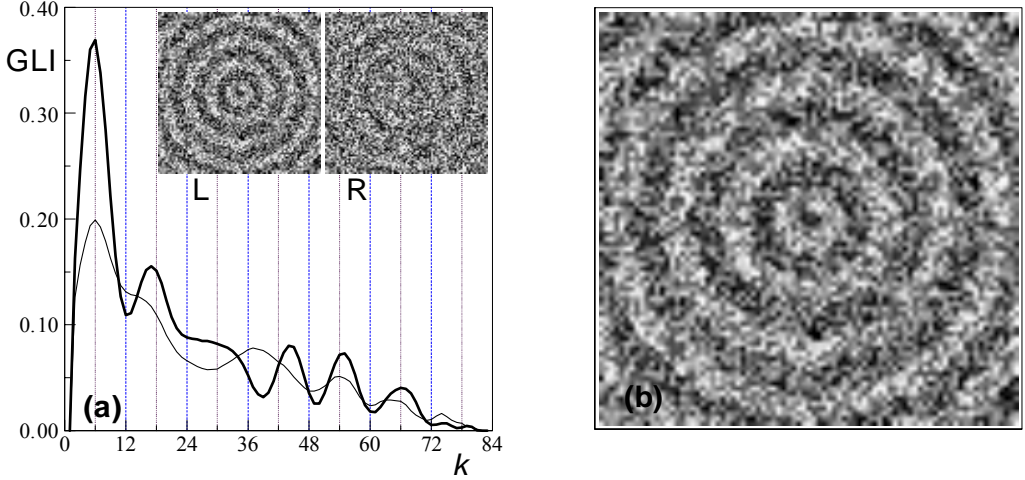

**Fig. 1.** (a) The grey level inhomogeneity (GLI) quantified by the generalized entropic measure $S_{gr,\Delta}(k; L)$, thick line, and $S_{gr,\Delta}(k; R)$, thin line, for two simulated $83 \times 83$ patterns, see the insets. The clearly seen effect of multiple intersecting curves (MIC) reveals the complex dependence of GLI on length scale. (b) The auxiliary $468 \times 468$ pattern composed of the sampled but non-overlapping $6 \times 6$ cells reproduces well the general structure (symmetry properties) of the initial L-pattern at scale $k_1 = 6$ referring to the first maximum.

Notice, that there is no periodicity for grey level values arrangements. Unequal inter-distances between the all-successive minimums confirm this. As a quite common feature one can observe along all scales the smaller variability of amplitude of GLI for the less contrasted R-pattern and visually more homogeneous. Such a feeling may be in part mistaken since it does not distinguish any particular length scale that is needed to define precisely, what is the value of the grey level inhomogeneity per cell. Our method quantifies this notion at each integer length scale for the actual macrostate AM($k$) with respect to the reference macrostate RM($k$) that describes the most uniform distribution of grey level values per cell. Taking into account the lack of periodicity and the nearly equal areas of grey level darker (lighter) subdomains one can envisage that subtle fluctuations in collection of local sums $G_i(k)$ may cause the surprising effect of multiple intersecting curves (MIC) first observed at scales $k = 11 \div 14$ in Fig. 1.

Another basic property of our measure is connected with the statistical grey level periodicity. In Fig. 2 we consider two simulated $360 \times 360$ chessboard patterns each consisting of 144 squares of size $30 \times 30$, see exemplary $90 \times 90$ parts in the insets.

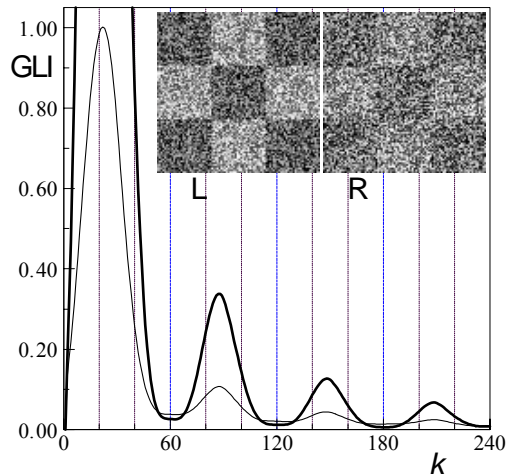

**Fig. 2.** Same as in Fig. 1 but for two simulated $360 \times 360$ patterns with statistical periodicity, see exemplary $90 \times 90$ parts in the insets; L-thick and R-thin lines. The weaker effect of MIC still occurs around positions of minimums.

The darker (lighter) squares were randomly contrasted to obtain the same as previously subdomain average of grey levels and same differences $\delta(L)$ and $\delta(R)$. The total averages of grey levels equal exactly to $\varphi(L) = \varphi(R) = 127.5$. The first peak for the curves of GLI(L), thick solid line, and GLI(R), thin solid line, is located at scale $k_1 = 22$. For L-pattern the maximum equals to about 3.77 but it is not shown for better visualization of multiple intersections at larger scales. From the construction of the patterns their statistical periodicity is expected at each multiple scale close to $k = 60$. This is in agreement with Fig. 2 even for much less contrasted R-pattern. Along all scales one can observe regular changes in the amplitude of GLI curves. Although much weaker the effect of MIC is still present within the considered range of scales. It should be stressed that the exactly equal areas of grey level darker (lighter) subdomains are used in the patterns of same symmetry properties.

Now, we focus on a hidden statistical grey level periodicity using as a starting point another binary $360 \times 360$ pattern, cf. Fig. 2(b) in Ref. [8]. It was subdivided into $30 \times 30$ squares and consisted of compact clusters (each of 120 black pixels) randomly placed one per cell. Now, the pattern was randomly contrasted to get for darker clusters (lighter surroundings) the averages of grey levels 75.5 (135.5) for L and 101.5 (131.5) for R-pattern, see Fig. 3 and the exemplary $90 \times 90$ insets. Similarly to L-case, another binary $360 \times 360$ but regular pattern of structure of a square lattice (SL), cf. Fig. 2(c) in Ref. [8], has been converted (it is not shown here). For this pattern each of the similar clusters occupies the centre of lattice cell. We have the same as previously $\delta(L)$, $\delta(R)$, $\delta(SL)$ and exactly equal $\varphi(L) = \varphi(R) = \varphi(SL) = 127.5$.

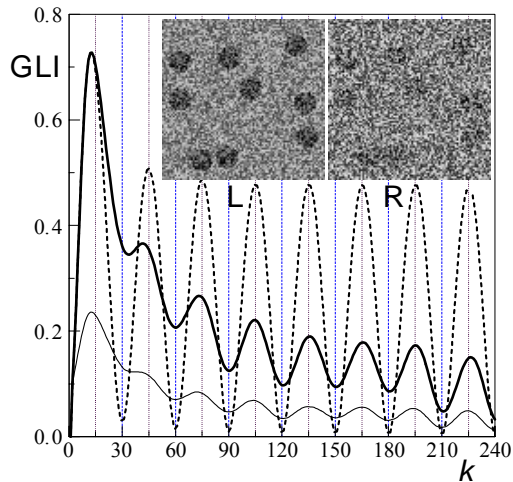

**Fig. 3.** Same as in Fig. 2 but for three simulated $360 \times 360$ patterns, two of them with deeply hidden statistical periodicity, see exemplary parts $90 \times 90$ in the insets; L-thick and R-thin solid lines. The third one relates to the L-contrasted regular pattern of structure of square lattice with the clusters occupying the centres of lattice cells, SL-dashed line (the R-contrasted corresponding regular SR-pattern is not shown here for clarity of the picture, see Tab. 1A in Appendix). Now, the effect of MIC is missing for L-R pair, but it is still present for L-SL and R-SL pairs. This shows that the effect of MIC can be easily caused by dissimilar symmetry properties of patterns.

For the L and R-patterns the deeply hidden statistical periodicity is not visible at first sight. However, in agreement with the construction of the patterns, all GLI curves oscillate quite regularly clearly showing the occurrence of a periodicity for each multiple scale $k = 30$. For L and R-curves the first maximum is located at $k_1 = 13$ while $k_1(SL) = 12$ with a nice correspondence to linear size (12 in pixels) of the simulated clusters. As expected, the measure does not distinguishes the different symmetries of L and SL-patterns at scales $k < 12$. Presently, for L-P pair there is no effect of MIC. This behaviour differs compared to that in Figs. 1 and 2. However, for L-SL and R-SL pairs of patterns, which show different symmetry properties the MIC effect appears. Supposedly, there is no single factor responsible for the revealed differences. In contrast to the previous cases, now the darker and lighter segments of the L and R-patterns are considerably dissimilar in size, what in this case probably favours more stable behaviour of the values of local sums $G_i(k)$ at each length scale. On the other hand, for L-SL and P-SL pairs, i.e. for the patterns dissimilar in symmetry features, this factor alone is not able to cause the MIC effect. It seems, that due to the complex dependence of GLI on length scale, the effect of MIC cannot be reduced to a single deciding factor, see Tab. 1A in Appendix.

Finally, we consider real grey level patterns adapted from images of Swedish Vacuum Solar Telescope (SVST) replaced in 2002 with a larger Swedish Solar Telescope (SST) on La Palma. Those images were recorded in 1997 and are available at website of The Solar Physics Group at NASA's Marshall Space Flight Center, see http://solarscience.msfc.nasa.gov/feature1.shtml. One can find there a series of 100 images of solar surface granulation evolving in time. Granules are related to the convective zone and are small cellular features with a typical length scale of about 1000 km (1 Mm) and a lifetime of roughly 10 minutes. They cover the entire Sun except for those areas covered by sunspots. Bright areas correspond to hotter fluid rising up from the interior and darker lanes to colder sinking inward fluid. Figs. 4(a) and (b) show the initial (#1) and ending (#100) frames, each of size $200 \times 200$ in pixels with the plate scale 0.083 arcsec/pixel $\cong$ 60 km/pixel [12]. The frames are taken in succession 1, 5, 10,…, 95, 100 from the granulation movie.

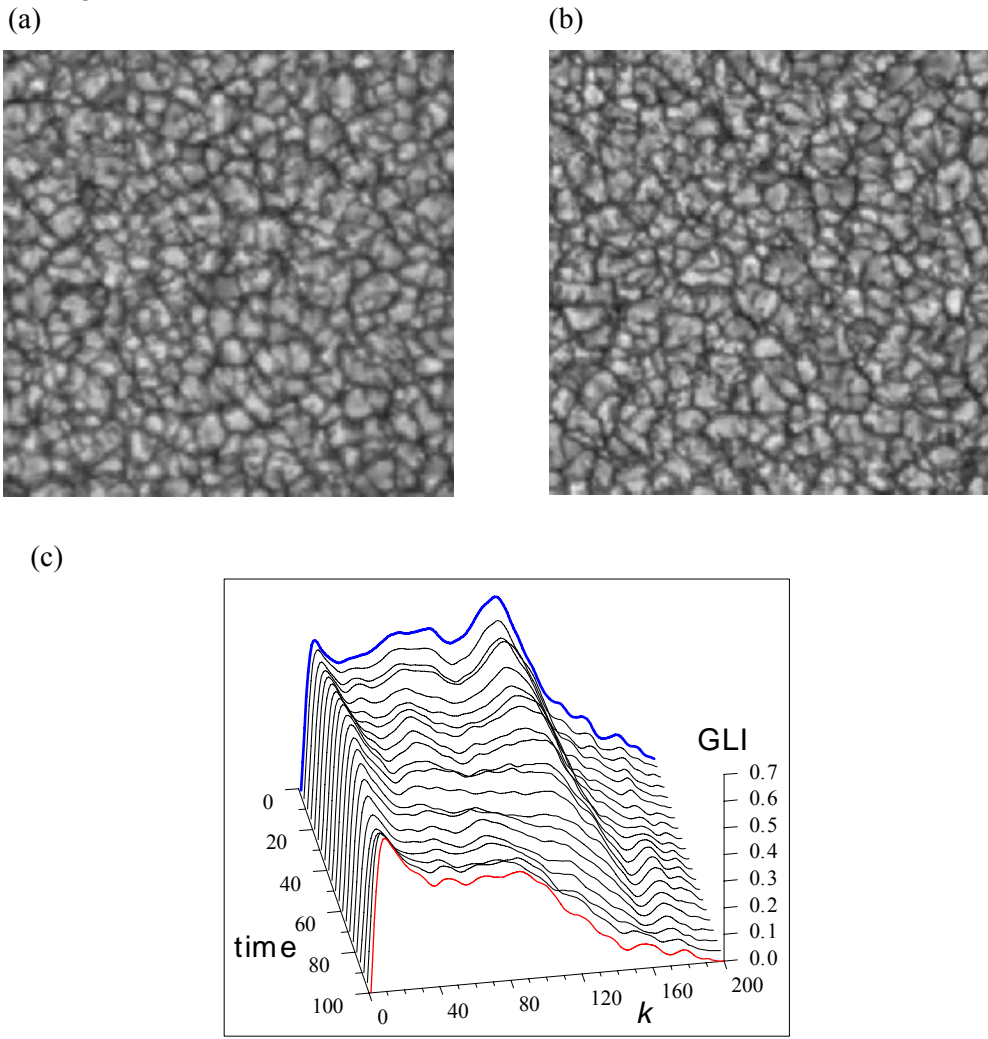

**Fig. 4.** (Colour online) similar as in Fig. 3 but for an adapted series (1, 5, 10,…, 95, 100) of solar granulation images recorded in 1997 as 470 kB MPEG movie, see http://solarscience.msfc.nasa.gov/feature1.shtml, from the Swedish Vacuum Solar Telescope that was replaced in 2002 with a larger Swedish Solar Telescope (SST). (a) and (b) The initial (#1) and ending (#100) frames each of size $200 \times 200$ in pixels; (c) Correspondingly, the thick (blue) and thin (red) lines. All curves show highly stable with time the position of the first peak at $k_1 = 9 \div 10$ in pixels corresponding to granulation ranging from 0.54 to 0.60 Mm. Initially, we observe a dominance of the third peak at $k \cong 110$ (6.6 Mm). This range may belong to mesoscale clusters, but we have not enough details about the data set. We can only say that at this scale the prevalent grouping of granules is a temporary process since along the time evolution the first peak recovers its dominant position.

Avoiding technical details, we would like to concentrate on basic properties of our measure, when it is employed to comparative multiscale analysis of a series of granulation patterns. In Fig. 4(c) the evolving multiscale 'surface' reveals some interesting features, which may be overlooked for a single pattern. Firstly, the stability with time of the first peak position is clearly seen, $k_1 = 9$ (for 16 frames) and $k_1 = 10$ (for 5 frames) in pixels. The length scales correspond to 0.54 and 0.60 Mm, respectively. Interestingly, the nearly constant with time the position of the first peak but around scale of 1 Mm has been reported in Ref. [13], where analysed photospheric intensity images of THEMIS telescope of the European Northern Observatory (Tenerife) recorded in 1999 and Dunn Solar Telescope (DST) of the National Solar Observatory (Sacramento Peak) recorded in 1996. However, instead of grey level patterns the authors used carefully prepared binary images and employed the normalized information entropy [5]. Secondly, at length scale $k \cong 110$ we detect the initially distinct third peak of our measure. During time evolution its domination decays and the first peak dominates again. Somewhat similar behaviour around of 8 Mm for DST96 data sets but for the time-averaged (10, 15, 20 min) normalized information entropy has been revealed in Ref. [14]. Again, the carefully binarized images were analysed but with a hexagonal sliding box. Such a box is more suitable in the case of granules than square moving window because of its symmetry corresponds closer to that of shape of granules.

Summarizing, we underline that the clearly seen for simulated patterns the effect of multiple intersecting curves reveals the complex dependence of GLI on length scales. This may be a good starting point for considering of construction of more universal, compared to proposal in [15], an entropic descriptor for a multiscale behaviour of spatially complex patterns. Among plenty papers devoted to entropy based complexity measures we mention two of them [16, 17], which are reviewing many important aspects of the notion of statistical complexity. This possibility deserves further research that will be reported elsewhere. Another future project relates to more advanced model including also unoccupied unit cells. They can be treated, for instance, as missing or incomplete data. It requires completely different combinatorial considerations and will be published elsewhere.

## 4. Conclusions

Despite its simplicity, the presented generalized entropic measure of grey level inhomogeneity is quite responsive for various details in pairs of simulated patterns of different types with the total average of grey level values equal (or very close) to 127.5. The rather intricate the effect of multiple intersecting curves, see Figs. 1-3 and Tab. 1A in Appendix, cannot be definitively reduced to a single deciding factor. It reveals the essential role of the length scales at which the grey level inhomogeneity is discussed. On the other hand, by the measure sensitivity we also mean its ability to detect hidden statistical grey level periodicity. Especially, for the low contrasted R-patterns when the darker segments are hardly seen. For evolving in time the photosphere granulation patterns the stability in time of the first peak position and temporary domination of the third one has been obtained by our GLI measure. The similar observations were reported by Consolini, Berrilli et al. (2003, 2005). However, they used normalized information entropy for carefully binarized granulation images of different data set. Regardless of statistical irregularities of the grey level patterns the examined measure shows its strength.

**Acknowledgements**

I thank Thomas Berger for sending me valuable information about granulation movie.

# Appendix

Tab. 1A  Collection of pairs of the simulated patterns with total average of grey level values equal (or very close) to 127.5 considered at length scales $1 \leq k \leq 83$ for Figs. 1 and 2, and $1 \leq k \leq 240$ for Fig. 3. The sign '+' ('−') denotes the occurrence (lack) of equal features in the paired patterns, for instance, '+' in the fourth column means the same differences of subdomain contrasting values, e.g., for (L-SL) pair is $\delta(L) = \delta(SL) = 60$ while '−' denotes different ones, e.g., for (R-SL) pair we have $\delta(R) = 30$ and $\delta(SL) = 60$.

| Figure # & pair | Symmetry properties | Darker/lighter subdomain areas | Darker/lighter contrasting | The effect of MIC |
|---|---|---|---|---|
| 1a (L-R)    | + | + | − | yes |
| 2  (L-R)    | + | + | − | yes |
| 3  (L-R)    | + | − | − | no  |
| 3  (L-SL)   | − | − | + | yes |
| 3  (R-SL)   | − | − | − | yes |
| 3  (L-SR*)  | − | − | − | no  |
| 3  (R-SR*)  | − | − | + | yes |
| 3  (SL-SR*) | + | − | − | yes |

\* For clarity of the picture, the SR-curve corresponding to regular R-contrasted SR-pattern is not shown in Fig. 3.